\documentclass[pre,amssymb,reprint,superscriptaddress,showpacs]{revtex4-1}
\usepackage{amsmath}
\usepackage{amsfonts}
\usepackage{amssymb}
\usepackage{epsfig}
\usepackage{graphicx}
\newcommand{\gdot}[0]{\dot{\gamma}}

\renewcommand{\phi}{\varphi}

\newcommand{\be}{\begin{equation}}
\newcommand{\ee}{\end{equation}}
\newcommand{\bea}{\begin{equnaray}}
\newcommand{\eea}{\end{equnaray}}
\newcommand{\ba}{\begin{align}}
\newcommand{\ea}{\end{align}}

\usepackage{color}

\begin{document}

\title{Diverging viscosity and 
soft granular rheology in non-Brownian suspensions}

\author{Takeshi Kawasaki}
\affiliation{Laboratoire Charles Coulomb, UMR 5221 CNRS, 
Montpellier, France}

\author{Daniele Coslovich}
\affiliation{Laboratoire Charles Coulomb, UMR 5221 CNRS, 
Montpellier, France}

\author{Atsushi Ikeda}
\affiliation{Fukui Institute for Fundamental Chemistry, 
Kyoto University, Kyoto, Japan}

\author{Ludovic Berthier}
\affiliation{Laboratoire Charles Coulomb, UMR 5221 CNRS, 
Montpellier, France}

\date{\today}

\begin{abstract}
We use large scale computer simulations and finite size scaling analysis
to study the shear rheology of dense three-dimensional 
suspensions of frictionless non-Brownian particles in the vicinity
of the jamming transition. We perform simulations of 
soft repulsive particles at constant shear rate,  
constant pressure, and finite system size and carefully study 
the asymptotic limits of large system sizes and infinitely  
hard particle repulsion. 
We first focus on the asymptotic behavior of the shear viscosity in the hard particle limit.
By measuring the viscosity increase over about 5 orders of magnitude, we are able to
confirm its asymptotic power law divergence close to the jamming transition.
However, a precise determination of the critical density and critical exponent is difficult due to the 
`multiscaling' behavior of the viscosity.
Additionally, finite-size scaling analysis 
suggests that this divergence is accompanied by a growing
correlation length scale, which also diverges algebraically. 
Finally, we study the effect of particles' softness and propose a natural extension of 
the standard granular rheology, which we test against our simulation data. Close
to the jamming transition, this ``soft granular rheology''
offers a detailed description of the non-linear rheology 
of soft particles, which differs from earlier empirical scaling forms.
\end{abstract}



\maketitle

\section{Introduction}

The jamming transition is widely studied in dense
disordered systems such as granular materials~\cite{jaeger_rmp},
emulsions~\cite{Emulsion}, suspensions of large colloids~\cite{wagner,Pouliquen}
and foams~\cite{Foam}.  A common feature to all these systems is that
thermal fluctuations play a negligible role on their dynamics,
i.e. they are non-Brownian, or `athermal'. Therefore the 
jamming transition is controlled by
density rather than temperature. Below the jamming packing fraction 
$\varphi_J$, the system flows with a finite viscosity when an
external force is applied. On approaching the jamming transition
from below, $\varphi \to \varphi_J$, the viscosity $\eta$ increases
dramatically and the system eventually develops a solid-like behavior
with a finite yield stress above $\varphi_J$. Despite recent efforts, a full
understanding of the rheological aspects of the 
jamming transition is still lacking, and this remains
an important challenge in soft condensed matter~\cite{rmp,vanhecke_review}.

The behavior of the viscosity of a non-Brownian suspension of 
infinitely hard particles not only serves as useful theoretical 
reference to understand the rheology of actual colloidal suspensions but 
is also relevant for hard granular particles~\cite{jaeger_rmp,bookyoel}. 
Empirically, it is believed 
that the viscosity of non-Brownian hard spheres exhibits 
a power law divergence on approaching $\varphi_J$:
\be \eta \approx  (\varphi_J-\varphi)^{-\beta},
\label{vis_power}
\ee
with an exponent $\beta \approx
2$~\cite{wagner,bonnoit,Pouliquen,teitel,andreotti,Claudin,Lerner}. 
A large number 
of empirical formulas have been proposed to describe the density 
dependence of $\eta$ for non-Brownian suspensions across a broad
range of density~\cite{wagner}, 
such as the Krieger-Dougherty form, $\eta \approx
(\phi_J - \phi)^{2.5 \phi_J}$~\cite{krieger_dougherty,krieger}. 
Here we focus on the asymptotic regime of large densities approaching
$\phi_J$ where essentially all formulas predict a power law divergence, 
as in Eq.~(\ref{vis_power}). Although both
experiments and simulations seem consistent with an algebraic 
divergence, several aspects remain unclear. Experimentally, it is
difficult to get accurate data over a broad dynamic range close to 
$\phi_J$, while frictional 
forces~\cite{silbert,makse,vanhecke,vanhecke_review,hayakawa_friction}
and flow localization~\cite{bonn,manneville} come as additional 
complicating issues. In computer simulations, interactions and 
flow geometries are typically well controlled, but the 
accessible viscosity range is usually quite modest, only about 2-3 
orders of magnitude. Also, a significant number of simulations were 
performed in 
two, rather than three, dimensions~\cite{teitel,andreotti}, 
paying little attention to finite size effects, which could
potentially emerge near an algebraic viscosity singularity.
Additionally, experiments and simulations have shown that for Brownian 
hard spheres, an apparent algebraic divergence of the shear viscosity,
detected from measurements obtained over a modest dynamic range, 
actually crosses over to a different functional form 
at larger density~\cite{gio}. The existence of a similar crossover 
has not been explored in non-Brownian systems, 
because it requires measurements over a larger 
dynamic range to establish (or disprove) the algebraic 
divergence in the absence of thermal fluctuations. 

Because they are soft objects, elastic particles can be compressed
above the jamming density $\varphi_J$. While this regime
cannot be accessed for hard grains, it is nevertheless
of experimental relevance for a large number of materials, 
such as emulsions and foams. In the jammed regime above 
$\phi_J$, the system is usually described by 
the empirical Herschel-Bulkley rheology combining a finite 
yield stress $\sigma_Y$ to shear-thinning behavior. Typically, the 
yield stress obeys the asymptotic  
relation~\cite{teitel,ikeda,ikeda_soft,OlssonTeitel2012,larson,hohler}
\be \sigma_Y \approx (\varphi-\varphi_J)^{\alpha},   
\ee 
where the critical exponent $\alpha$ depends
on the specific particle properties. 
Combining this relation with the asymptotic 
hard sphere behavior in Eq.~\eqref{vis_power}, Olsson and Teitel
suggested that the flow curves of frictionless soft particles can be collapsed 
on two master curves obtained by 
appropriately scaling the shear stress and the 
viscosity~\cite{teitel}. This scaling analysis 
indicates that the Newtonian viscous rheology below $\varphi_J$ and 
the soft particle rheology above $\varphi_J$ are in fact 
directly connected~\cite{OlssonTeitel2012}. The foundations of this
scaling analysis are largely empirical but physically appealing, as 
this allows to describe soft particles as `renormalized' hard 
spheres~\cite{OlssonTeitel2012,tom}. 
Note that since their original work,
Olsson and Teitel have thoroughly tested and revised their original
scaling approach, concluding in particular that 
important corrections to scaling are needed to describe the 
scaling properties of the shear viscosity~\cite{otherteitel2}.    
By revisiting the rheology of non-Brownian hard spheres, we can thus 
shed light on the soft particle rheology as well.  

In this work, we perform large scale simulations 
of the shear rheology in dense three-dimensional assemblies of soft harmonic 
particles. Our first goal is to considerably extend the dynamic range 
studied numerically to put the asymptotic divergence of the viscosity 
of non-Brownian suspensions of hard particles on firm grounds, 
paying special attention to finite size effects.  
Our second goal is to use our extended set of numerical data to 
carefully revisit the relationship between soft-
and hard-sphere rheologies, and propose a scaling description 
of the soft sphere rheology that is fully consistent with the hard sphere
limit. Our strategy is thus to perform simulations employing soft elastic
potentials~\cite{durian,ohern1,ohern2,hatano,teitel,ikeda,ikeda_soft,
OlssonTeitel2012}. 
The rationale is that, below $\varphi_J$, soft elastic particles
effectively behave as hard spheres in the limit of vanishing pressure
and shear stress. We can thus determine both hard and soft sphere rheologies
within the same numerical framework. 
The trade-off is that, although the computational effort of simulating soft
elastic particles is much lower than for hard particles, where 
overlaps are not allowed~\cite{wyart_hard_sheared}, this 
approach requires a careful asymptotic study of the hard sphere limit.

Using this approach, we are able to obtain viscosity measurements
in non-Brownian hard spheres in the large system 
size limit covering about 5 orders of magnitude. This allows
us to study the functional form of the viscosity on approaching $\phi_J$ 
with unprecedented accuracy. We confirm the 
algebraic nature of the viscosity divergence very near $\phi_J$, 
but our results also demonstrate that a precise determination of 
$\phi_J$ and the critical exponent $\beta$ is difficult due to the
inherent `multiscaling' nature of the granular rheology. Armed with 
these findings we then propose a simple extension of the hard
particle rheology to soft spheres, thereby suggesting a 
natural application of the granular rheology to soft systems.
While our scaling analysis is similar in spirit to the original proposal
by Olsson and Teitel~\cite{teitel}, we arrive to a different mathematical
model which suggests that the viscosity of soft particles does not 
obey a simple scaling form in the vicinity of the jamming transition.
Therefore, our approach gives novel physical insights into the 
emergence of strong corrections to scaling described  
in recent numerical work~\cite{otherteitel,otherteitel2}.

The organization of this paper is as follows.  In
Sec.~{\ref{sec:numeric}}, we describe our numerical methods.
In Sec.~{\ref{sec:rheology}} we present the results of
constant pressure simulations of soft
particles, and show how to reach the hard sphere limit. 
In Sec.~{\ref{sec:zero pressure}}, we perform a finite
size analysis of the data obtained in the hard sphere limit, which
allows us to describe the asymptotic behavior of the 
hard sphere system. In Sec.~{\ref{sec:finite pressure}} we
study the effect of particle softness and 
construct a soft granular rheology.
In Sec.~\ref{sec:summary}, we summarize our results and give some 
conclusions.

\section{Model and numerical methods}
\label{sec:numeric}

To investigate the flow behavior of non-Brownian particles over a wide range 
of flow rates, we performed overdamped Langevin dynamics simulations of 
a simple model suspension under shear flow at zero temperature.
Our model is a binary equimolar mixture of $N$ particles interacting via a 
harmonic potential~\cite{durian}. 
For two particles $i$ and $j$ having 
diameters $a_i$ and $a_j$, respectively, the harmonic potential reads
\be
U(r_{ij})=\frac{\epsilon}{2}\left(1-\frac{r_{ij}}{a_{ij}}\right)^2\Theta 
(a_{ij}-r_{ij}),
\label{pot}
\ee
where $r_{ij}$ is the distance between particles $i$ and $j$, 
$\epsilon$ is an 
energy scale,  $a_{ij}=(a_i+a_j)/2$, and $\Theta (x)$ is the 
Heaviside step function. 
In our simulations, the diameters of the small and large particles are 
$a$ and $1.4a$, respectively.
The particles evolve according to the following equation of motion   
\be 
\sum_j \frac{\partial U(|\vec{r}_{ij}|)}{\partial 
\vec{r}_{ij}} + \xi_s \left\{
\vec{v}_i(t)-\dot{\gamma}y_i(t)\vec{e}_y \right\} = \vec{0},
\label{em}
\ee
where $\vec{r}_{ij}=(x_{ij}, y_{ij}, z_{ij}) = (x_j-x_i,y_j-y_i,z_j-z_i)$, 
$\dot\gamma$ is the shear rate, and $\xi_s$ is a friction coefficient. 
We use Lees-Edwards periodic boundary conditions 
appropriate for homogeneous simple shear flows~\cite{allen}. 
From Eq.~(\ref{em}), we can define $\tau_0 = \xi_s a ^2 / \epsilon$ 
as the unit timescale, and $a$ as the unit lengthscale.  
Note that we use in this paper a very simple form for the energy 
dissipation, whose influence on the critical behavior has been 
debated in the literature. It was recently demonstrated that 
scaling behavior is actually not affected by the specific choice
of energy dissipation~\cite{recent_teitel}.  

Most of our simulations were carried out at 
constant pressure~\cite{Feller_ConstPressure,Kolb_ConstPressure}, 
in which the volume of the cell evolves according to 
\be 
\xi_V \dot{V}(t)-\{P-\hat{P}(t)\}=0,
\label{piston}
\ee
where $P$ is the prescribed pressure in units of $\epsilon/a^3$ 
and $\xi_V=10^{-4}\xi_s/a^4$. The latter value should be low enough to 
ensure stability of the pressure, but pressure fluctuations relax 
too slowly when $\xi_V$ is too small. 
The instantaneous value of the pressure, $\hat{P}(t)$, is defined as
\be 
\hat{P}(t)=\frac{1}{3V(t)}\sum_{j<i}\vec{r}_{ij}\cdot 
\frac{\partial U(|\vec{r}_{ij}|)}{\partial \vec{r}_{ij}}.
\ee 
In this setting, the control parameters of the simulation are 
the imposed pressure, $P$, the number of particles, $N$,
and the shear rate $\gdot$. 
For comparison, we also performed some constant volume 
simulations~\cite{ikeda}, see below.

An important observation is that pressure is expressed in units of 
the particle softness, $\epsilon/a^3$. This rather trivial remark implies 
that by reducing the pressure to zero, we probe configurations 
with smaller overlaps between the particles, so that 
the zero-pressure limit will effectively remove all 
overlaps between the particles if the density is below the jamming
density. Thus, {\it we recover the hard-sphere rheology by taking the 
$P \to 0$ limit.}  At finite pressure, particle softness
starts to play a role and we can then explore densities above the jamming 
transition.

We integrated the equations of motion in 
Eq.~(\ref{em}) using the method introduced in
Ref.~\cite{Kolb_ConstPressure} to perform constant pressure stochastic Langevin
simulations. This method is based on the ``reversible reference system
propagator algorithm''~\cite{RESPA}. The integration time step
was $\Delta t=1.0 \tau_0$, except for the smallest system ($N=100$) for which
we used $\Delta t=0.1 \tau_0$ instead, as explained below. 
For each studied state point, we let the system reach a steady state for a 
time $t_{\rm eq}$ and then accumulate measurements over a time $t_{\rm av}$.
The total simulation time $(t_{\rm av}+t_{\rm eq})$ is always larger than 
$1.0/\dot{\gamma}$ and we choose $t_{\rm av}=t_{\rm eq}$.
To investigate finite size effects, we performed simulations for 
six different system sizes: $N=10000$, 3000, 
1000, 500, 300, and 100. For each system size, we performed at least 
three independent runs to improve the statistics.
For the largest system sizes, we used a parallel version of the code 
which we run efficiently up to four cores. 

Two physical observables will be central in our analysis.
One is the $xy$-component of the instantaneous shear stress
\be \hat{\sigma}_{xy}(t) = - \frac{1}{V(t)}\sum_{j<i}
\frac{x_{ij} y_{ij}}{r_{ij}^2}\frac{\partial 
U(|r_{ij}|)}{\partial \vec{r}_{ij}}\cdot \vec{r}_{ij},
\ee
The other is the instantaneous volume fraction, which for our particular
binary mixture, reads:
\be
\hat{\varphi}(t)=\frac{N}{12V(t)} [a^3+ (1.4 a)^3].
\ee
The shear viscosity is then defined in terms of the average 
shear stress,  
$\sigma_{xy}
= (1/t_{\rm av})\int_{t_{\rm eq}}^{t_{\rm eq}+t_{\rm av}} \hat{\sigma}_{xy}(t)dt$,
\be 
\eta = \frac{\sigma_{xy} }{\dot{\gamma}}.
\label{visco}
\ee
We also define the average volume fraction 
$\varphi= (1/t_{\rm av})\int_{t_{\rm eq}}^{t_{\rm eq}+t_{\rm av}} \hat\varphi(t)dt $.
In the following, we will present stresses and viscosities expressed in 
units of $\sigma_0 = \epsilon /a^3$ and $\eta_0 = \xi_s/a$, respectively.
The influence of a solvent only appears in the 
viscous damping term in Eq.~(\ref{em}), and the solvent viscosity 
reads $\eta_s = \eta_0 / (3 \pi)$, which provides a reference 
value for the measured viscosity.

We optimized the simulation protocols and parameters so as to make 
the calculation of the viscosity in the hard sphere limit as efficient 
as possible. 
First, we optimized the discretization time step $\Delta t$. 
It must be large enough, so as to cut down the computational burden, 
but it must still produce physically sound results.
We compared the results for several physical properties, such as 
the packing fraction and the shear stress, 
using $\Delta t=0.5\tau_0$, $1.0\tau_0$, $1.2\tau_0$ for a 
broad range of shear rates at 
$P=10^{-5} \sigma_0$ and $N=10^4$. 
We found that the results for $\Delta t=0.5\tau_0$, $1.0\tau_0$, $1.2\tau_0$ 
were consistent within statistical uncertainties, while $\Delta t=1.5\tau_0$ 
led to numerically unstable results. Therefore we chose $\Delta t=1.0\tau_0$.
For the smallest system size ($N=100$) we used a time step $\Delta t=0.1\tau_0$
because we observed that   
simulations of such a small system were unstable with respect to lane 
formation close to the jamming point for larger time steps 
($\Delta t \geq 0.3\tau_0$). 

We also checked that the choice of the harmonic potential in
Eq.~\eqref{pot} allows the most rapid 
convergence to the hard sphere Newtonian limit. 
To test this, we considered a more general potential function 
of the form $U(r)=
\frac{\epsilon}{\alpha_r} (1-r/a)^{\alpha_r}$, 
and performed simulations with various $\alpha_r \in [1.2, 3]$. 
We found that a very soft potential (larger $\alpha_r$) allows to use a 
larger time step $\Delta t$ for time discretization, which 
speeds up simulations. However, for large $\alpha_r$, the hard sphere 
limit is reached for a smaller value of the shear rate $\gdot$, 
and the total simulation time becomes larger. By optimizing
$\Delta t$ for each value of $\alpha_r$, we found that the optimal
value of $\alpha_r$ is in the range $[1.75, 2.0]$, 
and we therefore decided to use the harmonic value, $\alpha_r = 2$. 
We emphasize that when the hard sphere limit is taken, 
the choice of the harmonic potential 
$\alpha_r = 2$ only appears as a numerical 
convenience, and the final results do not depend in any 
way on the value of $\alpha_r$. However, in Sec.~\ref{sec:finite pressure}
we discuss extensions of the hard sphere scaling laws to 
particles with a finite softness, and there the results 
depend quantitatively on the specific value of $\alpha_r$. 

\begin{figure}
\psfig{file=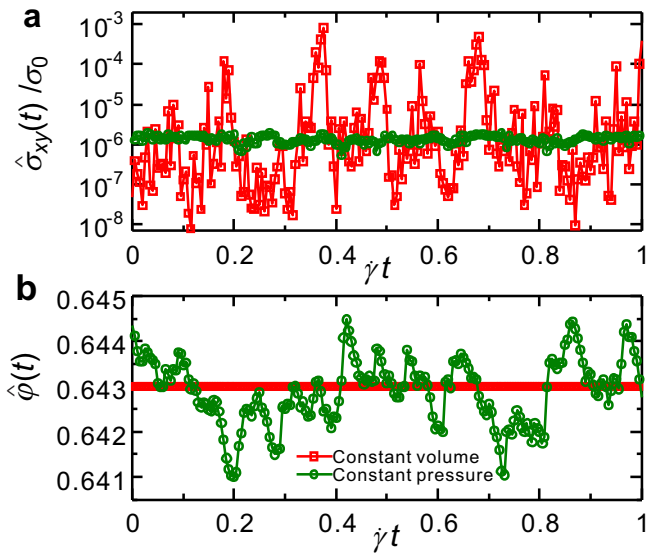,width=8.5cm}
\caption{\label{fig1} 
Time series of (a) the instantaneous shear stress, $\hat{\sigma}_{xy}(t)$, 
and (b) the instantaneous volume fraction $\hat{\varphi}(t)$, 
for $\gdot=3.98 \times 10^{-9}/ \tau_0$ and $N=1000$. 
Green circles represent the results of constant pressure 
simulations with $P=10^{-5} \sigma_0$, while the red squares 
represent those of constant volume simulations with 
$\phi=0.643$. Constant pressure simulations are superior as they 
yield small volume fraction fluctuations while 
drastically reducing the stress fluctuations.}
\end{figure}

Finally, we justify our choice to perform constant 
pressure simulations. The main motivation is 
to reduce the statistical uncertainty on the measurement of 
the physical observables of interest, which again 
results in more efficient simulations.   
In a finite system, physical observables may fluctuate differently
depending on the statistical ensemble.  In Fig.~\ref{fig1}-a and
Fig.~\ref{fig1}-b we show typical time series of $\hat\sigma_{xy}(t)$
and $\hat\varphi(t)$, respectively, during constant pressure and
constant volume simulations at a representative state point
($P=10^{-5} \sigma_0$, $\gdot=3.98\times 10^{-9}/\tau_0$, $\varphi=0.643$, and
$N=1000$).  This figure indicates that the fluctuations of
$\hat{\sigma}_{xy}(t)$ are considerably suppressed in constant
pressure simulations by comparison with constant volume simulations, 
while those of $\hat{\varphi}(t)$ remain
small enough.  More quantitatively, we
estimated the statistical uncertainty in the calculation of the
viscosity and volume fraction.  We divided the trajectory into 20 blocks
making sure the size of the block ($=t_{\rm av}$) was larger than the typical
correlation time and calculated the averages separately for each
block. Then we evaluated the standard deviation of these
block-averaged viscosities and densities.  We found that the
standard deviation of the viscosity is 5 times smaller in constant
pressure simulations. On the other hand, the standard deviation of
volume fraction by the ensemble average of three independent runs 
is as small as $3 \times 10^{-4}$ at this state point, and it
tends to decrease when $\varphi \to \varphi_J$. The uncertainty on
$\varphi$ is small enough not to affect our scaling analysis (see
for instance Fig.~\ref{fig7} below).  
Thus, we conclude that constant pressure
simulations should be preferred for the purpose of an accurate
determination of the viscosity $\eta(\varphi)$ near the jamming transition.

\section{Granular rheology in zero-pressure limit}
\label{sec:rheology}

\subsection{Granular rheology}

Let us start by presenting the flow behavior 
of our model of non-Brownian particles at finite pressure $P$ 
and finite $N$. We measure the average shear stress $\sigma_{xy}$, 
average volume fraction $\varphi$, and the viscosity 
$\eta$, as a function of the following three control parameters: 
$(P, N, \gdot)$. 

\begin{figure}
\psfig{file=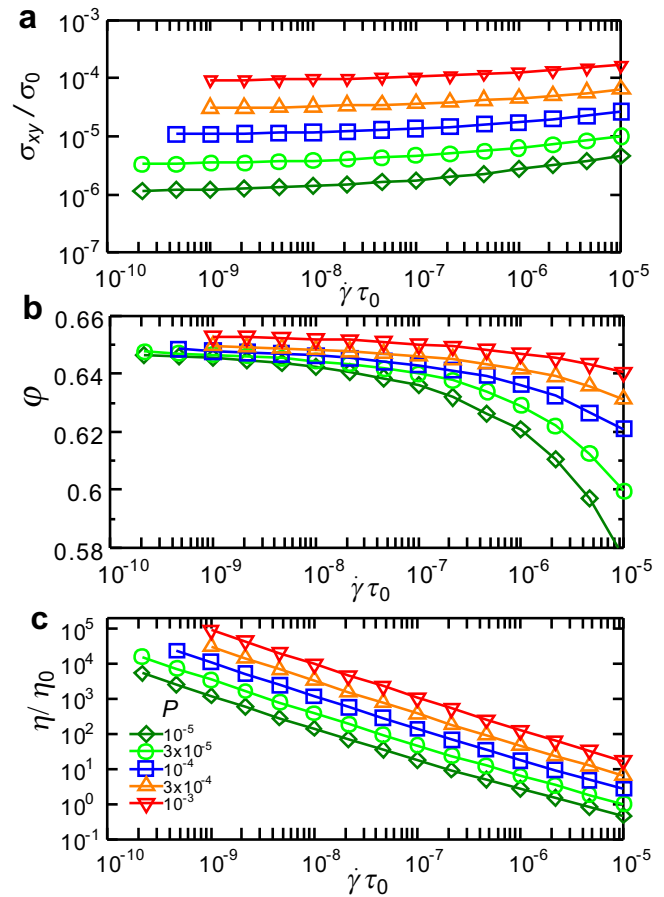,width=8.5cm}
\caption{\label{fig2} 
Evolution of our basic physical observables with the shear rate $\gdot$ 
in steady state constant pressure simulations at different $P$ 
values for $N=10^4$: (a) shear stress $\sigma_{xy}$, 
(b) packing fraction $\varphi$, (c) shear viscosity $\eta$.}
\end{figure}

The three panels in Fig.~\ref{fig2} present 
these three observables 
as a function of $\gdot$ at several pressures $P$ for a 
fixed system size of $N=10^4$ particles. 
As can be seen from Fig.~\ref{fig2}-a,  
$\sigma_{xy}$ increases moderately with increasing $\gdot$ 
at each given $P$, and it seems to converge to a finite value 
in the quasi-static limit $\gdot \to 0$ in each case. 
By decreasing the pressure, the shear stress 
shifts vertically towards smaller values, but the overall 
shape of the curves remains essentially the same.  

As shown in Fig.~\ref{fig2}-b, the packing fraction decreases with
increasing the shear rate in order for the pressure to remain
constant, which is nothing but the well-known dilatancy effect~\cite{bookyoel}. 
For a given shear rate,
the packing fraction increases weakly with increasing pressure due to
the softness of the potential, because particles overlap more when the
pressure increases. The packing fraction approaches a $P$- and
$N$-dependent limit as $\gdot \to 0$.  
At finite pressures this limiting packing fraction is above 
the jamming transition, so that
particles overlap at rest and store a finite pressure in the static packing.
In the $P \to 0, \gdot \to 0$ limit, 
the overlap between the particles must vanish and 
the volume fraction converges to the jamming transition point, $\phi_J$.
The system size dependence is discussed in Sec.~\ref{sec:zero pressure} below. 

Finally, let us focus on the behavior of the viscosity $\eta$ 
in Fig.~\ref{fig2}-c. We find that the viscosity decreases 
rapidly with $\gdot$ at constant pressure, mainly because 
the volume fraction also does.  
Clearly, the viscosity shifts systematically towards larger values 
as the pressure $P$ increases, because the density also increases.  

It is important to notice that 
for the soft particle system under study, it is very difficult to 
distinguish between Newtonian and shear-thinning regimes from the 
data set presented in Fig.~\ref{fig2}, because the viscosity is never
a constant, contrary to more traditional measurements employing 
constant volume techniques~\cite{teitel,ikeda,ikeda_soft}. 
When density is constant, the rheology takes a different form 
below jamming (where it has Newtonian and shear-thinning behavior), 
and above jamming (where it has yield and shear-thinning behavior), 
and the pressure exhibits complex density and shear rate 
dependences~\cite{teitel}.  
When the pressure is constant these two regimes do not 
appear separately, as shown in Fig.~\ref{fig2}. It is important
to recall that although the two rheologies appear different, 
they are of course fully equivalent.

Since the total simulation time increases inversely with $\gdot$, the
results in Fig.~\ref{fig2}-c indicate that simulations at small $P$
will achieve the same value of $\eta$ in a larger computational time
than those at large $P$.  However, as mentioned above, simulations at
large $P$ are also more likely to be influenced by the softness of the
potential, and presumably lie in the shear-thinning regime, as 
confirmed below. Because one
of our main goals is to extract a wide range of viscosity data in the
zero-pressure limit from finite pressure data, we will
carefully analyze the role of particle softness (or,
equivalently, finite pressures) in our results. Specifically, we must
find our way between the following two constraints: on the one hand, low
enough pressures are needed to attain the hard sphere limit; on the
other hand, large pressures are more efficiently simulated. Thus, we 
will seek data which satisfy the hard sphere limit with the largest
possible pressure.

\begin{figure}
\psfig{file=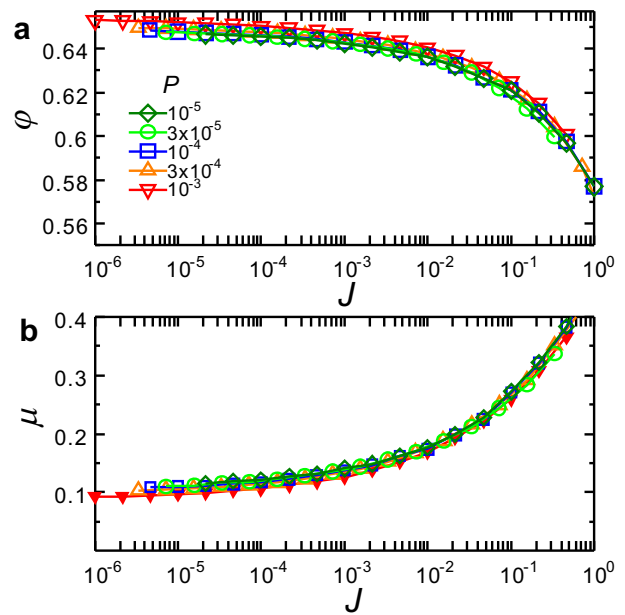,width=8.cm,clip}
\caption{\label{fig3}
The shear stress and volume fraction from 
Fig.~\ref{fig2} are represented in a dimensionless 
form appropriate for non-Brownian hard spheres, 
Eqs.~(\ref{defJ}, \ref{defmu}). The symbols are as in Fig.~\ref{fig2}.  
Deviations from full collapse originating from particle softness 
can be seen at large pressure, but are difficult to detect in this 
representation.}
\end{figure}

To analyze these data quantitatively, it is useful to use the zero pressure
hard sphere limit as a reference point. 
In this limit, the rheology simplifies drastically, because 
the system does not contain any energy scale. Therefore, 
the imposed pressure determines simultaneously the appropriate 
stress and time scales governing the behavior of the 
system~\cite{bookyoel,GDRMiDi,chevoir}. As is well-known in the literature of 
granular materials, this suggests to introduce dimensionless 
rheological quantities, and to express the shear stress and 
shear rates in dimensionless forms, since the packing fraction is 
already a non-dimensional quantity. Following the usual
notations~\cite{bookyoel}, we define the friction coefficient
as a dimensionless shear stress scale 
\be
\mu = \frac{\sigma_{xy}}{P},
\ee
and the viscous number as a dimensionless shear rate scale, 
\be
J= \frac{\gdot \eta_0}{P}.
\label{defJ}
\ee
In the zero pressure limit, we expect therefore that
the rheology is expressed through two simple 
relations~\cite{bookyoel}, 
\be
\mu = \mu(J), \quad \phi = \phi(J),
\label{defmu}
\ee
from which the shear viscosity is directly deduced as: 
\be
\frac{\eta}{\eta_0}  = \frac{\mu(J)}{J}.
\label{defeta}
\ee
An important conclusion is that since the relation $\phi(J)$ can be 
inverted, $J = J(\phi)$, the shear viscosity becomes 
a unique function of the volume fraction, 
$\eta(\phi)/\eta_0 = \mu ( J(\phi) ) / J(\phi)$, and in particular it does not 
depend on the imposed shear rate. In this limit, the 
rheology of the suspension is therefore purely Newtonian.

In Fig.~\ref{fig3}, we use these rescaled variables for our 
finite pressure simulations, and find that an essential part of the 
pressure dependence is scaled out by this scaled representation
of the data, compare with Fig.~\ref{fig2}.
Although deviations from this scaling can be seen, for instance,
for the largest pressure value, the data collapse look overall quite good. 
This is somewhat surprising
as soft repulsive particles display strong shear-thinning
effects near the jamming transition, as we discuss below. 
We are led to conclude that this presentation of the data
is actually a poor test of the influence of the particle softness
(or other perturbations to the hard sphere interaction), in particular
when statistical noise becomes significant, as is inherent for instance
to experimental measurements. As we show below, a
different representation of the data is therefore recommended. 

\subsection{Taking the zero-pressure limit}

Although the data collapse in Fig.~\ref{fig3} looks good, it is difficult 
to see how the data deviate from the hard-sphere behavior.
In order to observe and quantify these deviations more precisely, 
we study, for each value of $J$, the maximal pressure 
above which the data for $\mu(P,J,N)$ and $\phi(P,J,N)$ 
start to deviate from the $P \to 0$ limit, for a given
system size $N$. 
We then fit the $P$-independent parts of the data  
to the following functional forms~\cite{bookyoel,GDRMiDi,Lespiat,roux,Lerner}:
\begin{eqnarray}
\varphi(P \to 0,J,N) & = \varphi_J(N)-C_{\varphi}(N)J^{b_{\varphi}},
\label{phi_visco_no} \\
\mu(P \to 0,J,N) & = \mu_J (N) + C_\mu(N) J^{b_{\mu}},
\label{mu_visco_no}
\end{eqnarray}
where we explicitly included a system size dependence, which is 
the subject of Sec.~\ref{sec:zero pressure}.

\begin{figure}
\psfig{file=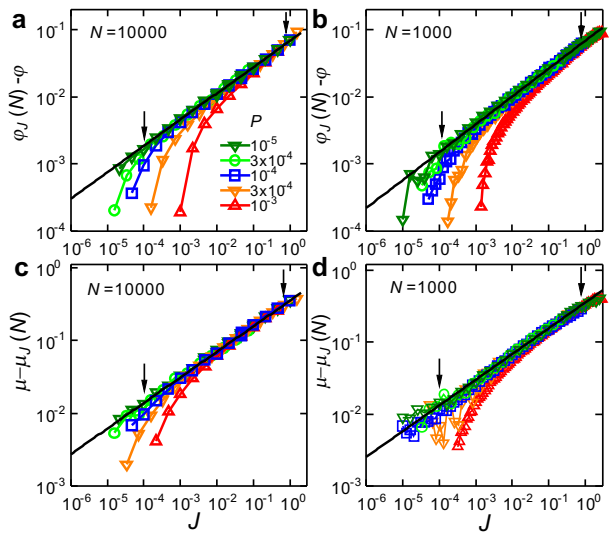,width=8cm,clip}
\caption{\label{fig4} 
Log-log representation of the power law
convergence of the friction coefficient and volume fraction
to their static values as the viscous number $J$ is decreased.  
The power laws in Eq.~(\ref{phi_visco_no}, \ref{mu_visco_no}) appear 
as black straight lines. In each panel, black arrows delimit the range 
of data where the zero-pressure limit is measured, and used for fits
to Eqs.~(\ref{phi_visco_no}, \ref{mu_visco_no}). Data are 
presented for two different system sizes.} 
\end{figure}

The results of this zero-pressure analysis are presented in 
Fig.~\ref{fig4} for two system sizes, $N=10^4$ and $N=10^3$. We 
show the $J$-dependence of 
$\phi_J(N) - \phi(P,J,N)$ and $\mu_J(N) - \mu(P,J,N)$ 
in a log-log scale. 
This amounts to representing the (nearly) collapsed data
in Fig.~\ref{fig3} using a logarithmic rather than linear 
representation of the vertical axis. In this different 
representation, finite pressure deviations are systematic and
become much easier to detect both for $\mu$ and for $\phi$. 

From the figure we see that deviations arise for 
$J$ lower than a crossover value that vanishes as $P \to 0$,
by construction. 
In Fig.~\ref{fig4}, we delimit by two vertical arrows
the range of viscous numbers $J$ where the $P=0$ limit is 
actually reached within the statistical accuracy of the data, 
and where the `envelope' of the converged data can be analyzed. 
By fitting this envelope to the functional forms given in 
Eqs.~(\ref{phi_visco_no}, \ref{mu_visco_no}), we deduce
for each system size $N$ the fitting parameters involved
in these expressions. The result of these fit are indicated by 
full lines in Fig.~\ref{fig4}. They describe a power law 
convergence of $\phi$ and $\mu$ towards their $J \to 0$ asymptotic 
values. We emphasize that these power laws are obeyed over a significant
range of inertial numbers $J$ of about 4 decades, and we find 
that these power laws hold independently for each different system 
size studied numerically, $N=10^2-10^4$.     

\section{Large-$N$ limit of granular rheology}
\label{sec:zero pressure}

\subsection{`Brute-force' analysis of numerical data}
\label{brute}

Having dealt with the zero-pressure limit, we now proceed to  evaluate 
quantitatively the $N\to \infty$ limit, so as to get rid of finite size 
effects. Finite size effects have been discussed in earlier simulations 
of non-Brownian particle 
systems~\cite{roux,ohern1,ohern2,pinaki,recent_teitel,otherteitel,otherteitel2}.  

\begin{figure}
\psfig{file=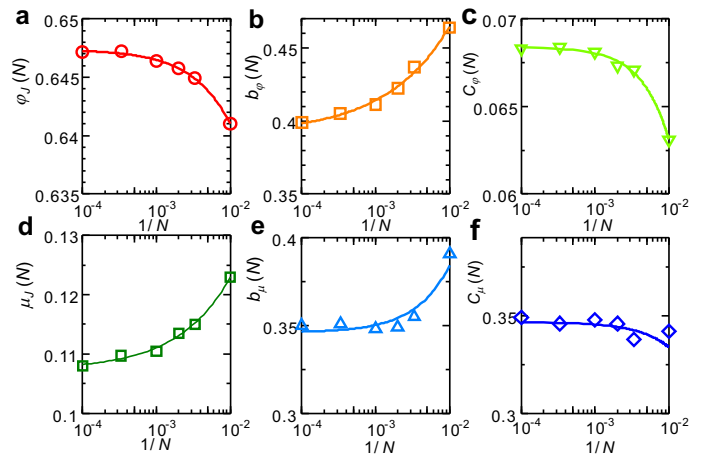,width=9cm,clip}
\caption{ \label{fig5} 
System size dependence of the fitting parameters 
appearing in Eqs.~(\ref{phi_visco_no}, \ref{mu_visco_no}). 
The lines are fits describing a power law approach to the asymptotic 
$N \to \infty$ limit, as in Eq.~(\ref{O_finite}), with values 
reported in Table~\ref{tb:table2}.}
\end{figure}

In Fig.~\ref{fig5} we show the $N$-dependence of the parameters 
$\varphi_J(P \to 0,N)$, $C_\varphi(N)$, $b_{\varphi}(N)$, $\mu_J(P \to 0,N)$, 
$C_\mu(N)$, and  $b_{\mu}(N)$, defined in Eqs.~(\ref{phi_visco_no},
\ref{mu_visco_no}), and obtained from independently fitting the data shown 
Fig.~\ref{fig4} for different system sizes. The fact 
that these parameters evolve smoothly with system size 
demonstrates the high quality of our numerical data, since these  
parameters are measured from statistically independent sets of data.

We find empirically that the system size dependence 
of these parameters is also well described by power laws
\be
{\cal O} (N)= {\cal O} (N \to \infty) +
{A_{\cal O}}{N^{-\alpha_{\cal O}}},
\label{O_finite}
\ee
where ${\cal O}(N)$ stands for any of the observables under study
measured for a finite $N$, ${\cal O} (N \to \infty)$
its fitted asymptotic value assuming a power law convergence with $N$
with an exponent $\alpha_{\cal O}$.  

It is not surprising that such power laws are obeyed because
these parameters change very little when the total number of particles
is increased by two orders of magnitude. These power laws can thus
be seen at this stage as a convenient empirical 
method to determine quantitatively
the asymptotic value of the parameters entering the granular 
rheology. The extrapolation to infinite system size  
eventually allows us to obtain the asymptotic behavior of the system 
as both the hard sphere (zero-pressure) and thermodynamic limits are taken, 
$N \to \infty, P \to 0$.  
The full set of fitting parameters used in Fig.~\ref{fig5} are summarized 
in Table~\ref{tb:table2}.

\begin{table}[b]
  \begin{tabular}{| c | cccccc |} \hline
${\cal O}(N)$ & $\phi_J$ & $b_\phi$ & $C_\phi$ & $\mu_J$ & 
$b_\mu$ & $C_\mu$  \\ \hline
${\cal O}(N \to \infty)$ & 0.6474 & 0.391 & 0.0683 & 0.108 & 0.346 
& 0.347  \\  
$A_{\cal O}$ & -0.323 & 0.756 & -1.15 & 0.279 & 2.93 & -1.44 \\
$\alpha_{\cal O} $ & 0.852 & 0.505 & 1.17 & 0.626 & 0.946 & 1.02 \\ \hline 
\end{tabular}
\caption{Sets of parameters obtained by fitting the system size dependence
of several physical observables 
with the functional form in Eq.~(\ref{O_finite}).}
\label{tb:table2}
\end{table}

In particular, our data indicate that in the double limit 
$N \to \infty, P \to 0$, the asymptotic granular rheology of 
the present binary system reads:   
\be
\phi = \phi_J - C_\phi J ^{b_\phi}, \quad
\mu = \mu_J + C_\mu J^{b_\mu}, \label{asymptotic}
\ee
with the estimated asymptotic values:
\begin{eqnarray} 
\phi_J & = 0.6474, \quad  b_\phi = 0.391, \\
\mu_J  & = 0.108, \quad b_\mu = 0.346.
\end{eqnarray}

\subsection{Scaling analysis: Diverging correlation length}
\label{sec:finite size limit}

\begin{figure}
\psfig{file=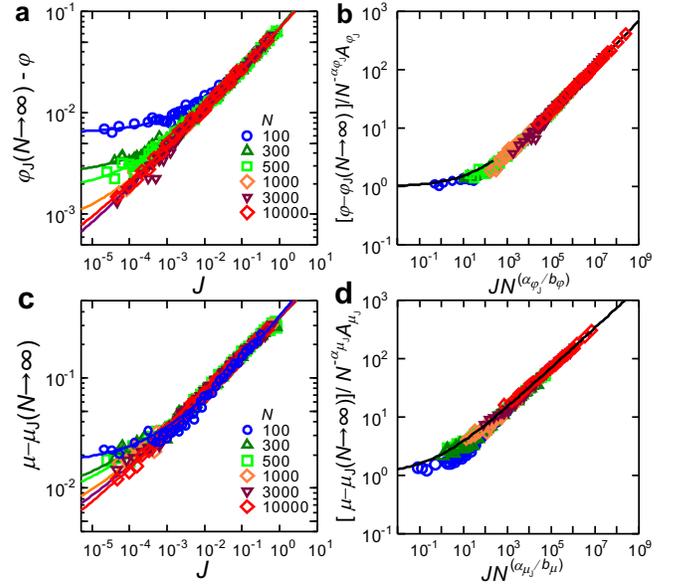,width=8.5cm,clip}
\caption{\label{fig6} 
Scaling analysis of finite size effects.
(a) Emergence of finite size effects for the volume fraction $\phi$,
and (c) for the friction coefficient $\mu$.
The solid lines represent fits using Eqs.~(\ref{O_finite}, \ref{asymptotic}).
(b) Scaling plot for the volume fraction using Eq.~(\ref{scaling_phi}).
(d) Scaling plot for the friction coefficient using Eq.~(\ref{scaling_mu}).
Solid lines in (b, d) use parameters in Table~\ref{tb:table2}. }
\end{figure}

It is fruitful to revisit the above extrapolation 
to infinite system sizes using a somewhat simpler, but physically
more illuminating, scaling analysis. 
Using the results of the above `brute force' fitting procedure to extract the 
large-$N$ limit of the results, we can represent in Fig.~\ref{fig6}-a 
the dependence of $\varphi_J(N \to \infty)-\varphi$ and in Fig.~\ref{fig6}-b
the dependence of $\mu_J(N \to \infty)-\mu$ for the data 
measured in the zero-pressure limit. It is important to note
that we use the infinite system size asymptotic  
values $\phi_J(N \to \infty)$ and $\mu_J(N \to \infty)$ to 
represent data at finite $N$, an approach which 
reveals finite size effects more clearly. 
Recall also that these measurements
are obtained in the zero-pressure limit, as we are only concerned 
with finite size effects in the present subsection.

In the different representation of Fig.~\ref{fig6}, finite size corrections to 
the asymptotic behavior can be better appreciated and they seem to 
take a particularly simple and suggestive 
form. For a given system size, the data  exhibit 
two distinct regimes. For large $J$, the data show little 
size dependence, while at small $J$ clear deviations 
are observed. Additionally, the crossover viscous number separating the two
regimes strongly depends on $N$, the deviations shifting to smaller
 $J$ values as $N$ increases and disappearing 
as $N \to \infty$, by construction.

This qualitative description suggests that finite size effects 
can be described analytically using the following scaling form:
\begin{eqnarray}
\phi(J,N) & = [ \phi_J(N \to \infty) + A_{\phi_J} N^{-\alpha_{\phi_J}} ] - C_\phi 
J^{b_\phi}, \label{scalingphi} \\
\mu(J,N) &  = [ \mu_J(N \to \infty) + A_{\mu_J} N^{-\alpha_{\mu_J}} ] + C_\mu 
J^{b_\mu}, \label{scalingmu}
\end{eqnarray}
which amounts to neglecting the system size dependence of the prefactors 
$C_\phi$ and $C_\mu$ and the exponents $b_\phi$ and $b_\mu$ in the 
`brute force' description of Sec.~\ref{brute}. 
The physical content of Eqs.~({\ref{scalingphi}, \ref{scalingmu}) is that 
a finite size system obeys the asymptotic granular rheology 
of Eqs.~(\ref{asymptotic}), but with values for the 
jamming density $\phi_J$ and the friction coefficient $\mu_J$
which are `renormalized' by finite size corrections. 
In particular, this implies that smaller systems jam at a 
smaller density, Fig.~\ref{fig5}-a, with a larger 
value of the friction coefficient, Fig.~\ref{fig5}-b.
This description is fully consistent with earlier work~\cite{ohern1,roux}.  

This simplification then suggests that the finite size data 
can be collapsed using the following representation:
\begin{eqnarray}
\frac{\phi(J,N) - \phi_J(N \to \infty)}{A_{\phi_J} N^{-\alpha_{\phi_J}}} & 
= {\cal F}_\phi 
(J N^{\alpha_{\phi_J}/b_\phi}),  \label{scaling_phi} \\
\frac{\mu(J,N) - \mu_J(N \to \infty)}{A_{\mu_J} N^{-\alpha_{\mu_J}}} & 
= {\cal F}_\mu 
(J N^{\alpha_{\mu_J}/b_\mu}), \label{scaling_mu}
\end{eqnarray}
where ${\cal F}_\phi(x)$ and  ${\cal F}_\mu(x)$ are two scaling functions
with the asymptotic behavior 
${\cal F}_{\phi,\mu} (x \to 0) = 1$ (when finite size effects dominate), and 
${\cal F}_{\phi,\mu} (x \to \infty) \sim x^{b_{\phi,\mu}}$
(when finite size effects are absent). 

The scaling plots resulting from Eqs.~(\ref{scaling_phi}, \ref{scaling_mu}) 
for $\varphi$ and $\mu$ are shown 
in Figs.~\ref{fig6}-b and Fig.~\ref{fig6}-d, respectively. 
We can see that our scaling hypothesis describes the data 
in a satisfactory manner. The full lines in the figures
are from the fitted values obtained above 
in Sec.~\ref{brute}, see Table~\ref{tb:table2}, and provide
an acceptable analytical description of the numerical results
over a broad range of scaled variables. 

Remarkably, this data collapse suggests that the approach 
to the {\it jamming transition as $J \to 0$ in non-Brownian hard particles
under shear is accompanied by a diverging length scale}, $\xi(J \to 0) 
\to \infty$. 
From the above scaling, we deduced that for each system size, 
there exists a crossover $J$-value below which finite size effects are observed.
This can be interpreted by saying that finite size effects are observed
whenever $\xi(J)$ becomes comparable to the linear size of 
the system, $\xi \sim N^{1/d}$, where $d$ is the space dimensionality.
Combining this argument to the scaling forms 
in Eqs.~(\ref{scaling_phi}, \ref{scaling_mu}) suggests that 
\be
\xi \sim J^{-\nu},
\label{nu}
\ee
with $\nu \approx b_\phi / (\alpha_{\phi_J} d) 
\approx b_\mu / (\alpha_{\mu_J} d)$.
From the values reported in Table \ref{tb:table2} we obtain 
$\nu \approx 0.153$ (from $\phi$) and $\nu \approx 0.184$ (from 
$\mu$). These numerical estimates are 
consistent with a unique value for $\nu$. 
Note that the reliability of this scaling analysis depends on the
validity of the assumption that both the pair of exponents $b_\mu$ and $b_\phi$
and the pair of coefficients $C_\mu$ and $C_\phi$ 
are independent of the system size $N$.
The data in Fig.~\ref{fig5} show 
that $b_\mu$ and $C_\mu$ are actually nearly independent of $N$, 
so that the value $\nu \approx 0.184$ is probably more 
reliable. 

To discuss the value of this exponent, it is useful to first
remember that once the zero-pressure limit has been taken, 
the jamming transition at $\phi_J$ is approached along 
a specific path, namely, 
as the rescaled shear rate goes to zero, $J \to 0$.  
This is therefore a different path in control parameters space 
from more traditional approaches where for instance 
static packings ($J=0$) are produced at densities
approaching $\phi_J$ from below~\cite{ohern2}, 
or where the shear rate is decreased
exactly at the jamming density~\cite{recent_teitel}.
In the present case, the packing fraction is actually also changing
along the way.   

We can nevertheless compare the obtained value for $\nu$ to a
number of literature results. First we notice that the rather small value of 
the exponent $\nu$ in Eq.~(\ref{nu}) is consistent 
with a recent discussion of the correlation length
characterizing the response to a local 
perturbation in a similar numerical model~\cite{during,degiuli},
where the value $\nu \approx 0.15$ is reported. 
A second set of approaches dealing with the rheology of dense 
amorphous materials has demonstrated the need to introduce a 
correlation length in dense flows to account for the 
emergence of `non-local' rheological effects~\cite{KEP,kamrin,bruno}. 
Despite the diversity of empirical models, they all introduce a 
lengthscale $\xi_{nl}$ 
which diverges at jamming as a square root of the distance
to yielding, $\xi_{nl} \sim (\mu - \mu_J)^{-1/2}$. Converting our findings
in this representation, we obtain $\xi \sim J^{-\nu} \sim (\mu - 
\mu_J)^{-\nu / b_\mu}$, where $\nu / b_\mu = 1/(\alpha_{\mu_J} d) \approx 0.53$.
The close agreement between our findings and 
these empirical models (adjusted to best fit numerical data)
suggests a possible deep
connection between the finite size effects we detect directly in our 
work, and non-local effects observed experimentally.   
To our knowledge, the correlation length exponent has not been 
directly determined experimentally~\cite{olivier}, but we note 
that the smallness of $\nu$ means that $\xi$ diverges 
relatively slowly as the jamming transition is approached.
This probably implies that it should be 
difficult to measure $\nu$ experimentally, unless the system is 
extremely close to jamming~\cite{Pouliquen,Lespiat}.  

\subsection{Asymptotic results for the shear viscosity}

Having obtained `converged' numerical results for the 
packing fractions and the friction
coefficient, i.e. data that satisfy both the hard sphere ($P \to 0$)
and the large system size limit ($N \to \infty$), we can now evaluate
the physical behavior of the shear viscosity in the limit 
where neither particle softness nor finite size effects play any role.

To this end, we combine the converged values of $\mu$ and $\phi$ 
and use Eq.~(\ref{defeta}) to obtain the shear viscosity.  
In Fig.~\ref{fig7}, we represent $\eta$ as a
function of $\varphi_J - \phi$ for several finite pressures and system
sizes, carefully collecting the data that satisfy 
both $P \to 0$ and $N \to \infty$
limits. Our numerical measurements cover about 5 orders of magnitude
in viscosity, which is significantly larger than the typical
range accessed experimentally and in earlier simulation 
studies~\cite{andreotti,ikeda,OlssonTeitel2012}.

\begin{figure}
\psfig{file=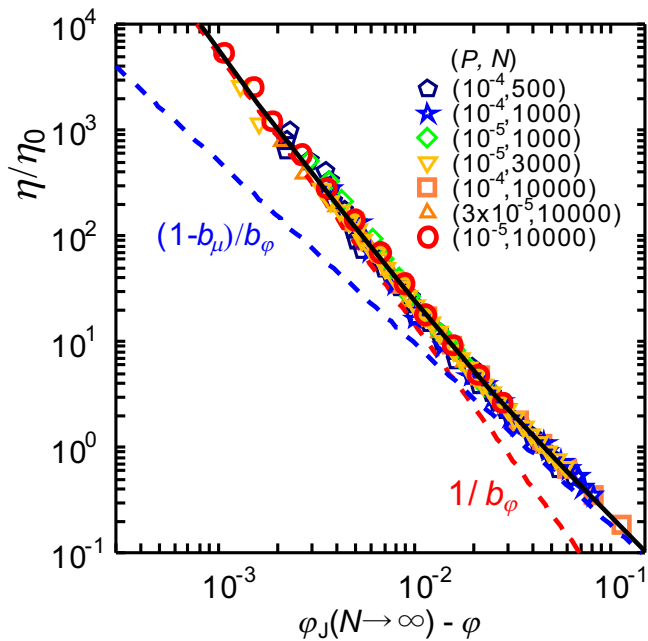,width=8.5cm,clip}
\caption{\label{fig7} 
Shear viscosity after the hard sphere ($P \to 0$) and
the large system size ($N \to \infty$) limits are taken.  
Symbols are measurements for different $N$ and $P$
which satisfy the $N \to \infty, P \to 0$ limits. 
The black line represents Eq.~(\ref{eta}) with parameters
taken from Table~\ref{tb:table2}. Red and blue dashed lines 
separately represent the two power law contributions in Eq.~(\ref{eta}).}
\end{figure}

Collecting the above results, 
we can easily get an analytical description of these 
viscosity data by combining  
Eq.~(\ref{asymptotic}) with Eq.~(\ref{defeta}),
so that the relationship between $\eta$ and $\varphi$ in the $P \to 0$ and 
$N\to\infty$ limits reads:
\be
\frac{\eta}{\eta_0}=\mu_J \left[\frac{C_\varphi}{\varphi_J 
-\phi}\right]^{\frac{1}{b_{\varphi}}}
+C_\mu\left[\frac{C_\varphi}{\varphi_J -\phi}\right]^{\frac{1-
b_{\mu}}{b_{\varphi}
}},
\label{eta}
\ee
where all parameters should be taken with their $N\to\infty$ values (we 
have omitted this limit in Eq.~(\ref{eta}) to simplify the notation).
Using the numerical values reported in Table~\ref{tb:table2}, we can 
directly compare the analytical expression in Eq.~(\ref{eta}) to the 
numerical data. This is shown as the the black line going through the 
symbols in Fig.~\ref{fig7}. The very good agreement confirms the consistency
of our data analysis.

We note that Eq.~\eqref{eta} differs from the single power law function,  
which was used in several previous 
studies~\cite{Pouliquen,teitel,andreotti,Claudin} to fit viscosity data,
but is mathematically consistent with the description 
of corrections to scaling in Ref.~\cite{otherteitel2}.  
The viscosity behaves in fact as the 
sum of two power laws with different exponents, such that 
taking the final limit of $J \to 0$ in our data, we obtain the 
asymptotic behavior where the largest of the two exponents 
dominate:
\be
\eta \sim (\phi_J - \phi)^{-1/b_\phi} \sim (\phi_J - \phi)^{-2.55}.
\ee
Our numerical analysis, which considerably expand previous work,
confirms the robustness of the algebraic divergence of the shear 
viscosity in non-Brownian hard particles near the jamming density.
In particular, we detect no sign of a crossover towards a sharper density
dependence, as observed in thermalized assemblies of hard 
particles~\cite{gio}. This different qualitative behavior between 
Brownian and non-Brownian suspensions confirms the distinct nature of 
glass and jamming transitions in colloidal assemblies~\cite{tom,ikeda,PZ}. 

In Fig.~\ref{fig7}, the two power law contributions
in Eq.~(\ref{eta}) are separately highlighted with dashed lines. 
It is clear that 
the shear viscosity crosses over from one power law divergence to 
another, the crossover between the two occurring relatively close to 
the jamming transition, for $\Delta \phi \approx 0.02$. This implies 
that a precise determination of the critical exponent governing 
the viscosity divergence is challenging, as the scaling regime starts 
at the lower boundary of the regime explored in the most recent 
experiments~\cite{Pouliquen}. This finding might also explain the 
diversity of critical exponents $\beta$ that have been reported in 
the literature, because a power law fit of a narrower range of 
data could yield an exponent lying anywhere 
in the range of $\beta \in 
[(1-b_{\mu})/b_{\varphi}, 1/b_{\varphi}] \approx [1.67, 2.55]$. 
This is broadly consistent with the rough accepted value 
$\beta \approx 2$ found in 
many previous works~\cite{Pouliquen,teitel,andreotti,Claudin}, 
but somewhat smaller than the recent prediction $\beta = 2.83$ for its 
numerical value~\cite{degiuli}.  

\section{Soft granular rheology} 
\label{sec:finite pressure}

\subsection{Scaling hypothesis for soft grains}

Because of the finite energy scale introduced by the particle
softness, the rheology of soft particles is not uniquely governed by
the viscous number $J$, and non-linear effects therefore come in, as described
in Fig.~\ref{fig4}. In this section, we shall study the effect of
particle softness or, equivalently, of finite pressures. We shall
present detailed results for two system sizes to establish the 
robustness of our analysis, but we do not explicitly
account for the finite size dependence of physical observables 
as we did for the zero-pressure limit. The main variable of interest
in this section is therefore the pressure $P$. Note that 
in this section, the specific choice of a harmonic repulsion between 
the particles becomes relevant. 

The finite pressure corrections to the granular rheology 
shown in the data of Fig.~\ref{fig4} are very suggestive, 
and qualitatively reminiscent of the finite size effects 
shown in Fig.~\ref{fig6}. Indeed these data show that for a given 
pressure value, the data at large $J$ are not affected by 
$P$, while deviations are seen at small enough $J$. Crucially, 
the crossover $J$-value separating the two regimes is 
pressure dependent, and vanishes, by construction, when $P \to 0$.  

Inspired by the scaling hypothesis performed in 
Sec.~\ref{sec:finite size limit}
to account for finite size effects, we make a similar 
hypothesis for finite pressure and assume that 
finite pressure corrections  can be analytically accounted 
for by generalizing 
the granular rheology in Eq.~(\ref{asymptotic}) to the following 
{\it `soft granular rheology'}:  
\begin{eqnarray} 
\varphi(P,J) & = \varphi_J(P)-C_{\varphi} J^{b_{\varphi}},
\label{phi_scale} \\
\mu(P,J) & = \mu_J(P)+C_\mu J^{b_{\mu}},
\label{mu_scale}
\end{eqnarray}
where $\phi$ and $\mu$ are measured at finite pressures $P$
and viscous numbers $J$. 
Comparing to Eq.~(\ref{asymptotic}), these expressions 
capture the idea that a finite pressure simply `renormalizes'
the hard sphere jamming density $\phi_J$ and friction coefficient 
$\mu_J$ to pressure dependent values, while the functional form 
of the rheology is unaffected.

\begin{figure}
\psfig{file=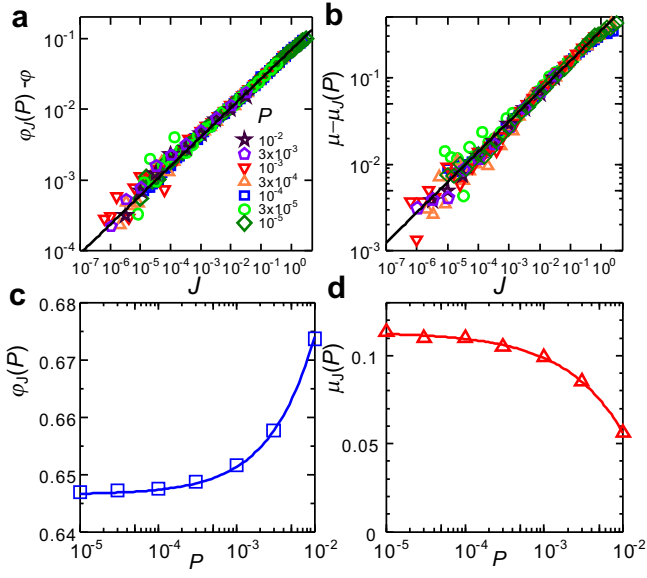,width=8.5cm,clip}
\caption{\label{fig8} 
Test of the soft granular rheology extending the 
hard sphere rheology to soft particles. Data are shown for $N=10^3$. 
(a) $\varphi_J(P) - \varphi$ vs. $J$ for several $P$ values;
the solid line represents Eq.~(\ref{phi_scale}).
(b) $\mu_J(P)- \mu$ vs. $J$ for several $P$ values; the
solid line represents Eq.~(\ref{mu_scale}).
(c) Pressure dependence of $\varphi_J(P)$; the
solid line represents Eq.~(\ref{phi_P}) with $x=0.75$.  
(d) Pressure dependence of $\mu_J(P)$; the 
solid line represents Eq.~(\ref{mu_P}) with $y=0.56$.}
\end{figure}

We test this very simple idea in Figs.~\ref{fig8}-a,b where we represent
$\varphi_J(P,J)-\varphi$ and $\mu(P,J)-\mu_J(P)$ as functions of 
the viscous number $J$ for different pressure values. 
To construct these figures, we determine for each pressure the
values of $\phi_J(P)$ and $\mu_J(P)$ that yield the best collapse of the
data. Clearly, this procedure removes the systematic pressure
dependencies observed in Fig.~\ref{fig4}, which shows that our
hypothesis in Eqs.~(\ref{phi_scale}, \ref{mu_scale}) is 
satisfied within the statistical accuracy of our data.

The measured functions $\phi_J(P)$ and $\mu_J(P)$ are 
reported in Figs.~\ref{fig8}-c,d. These data show that  
finite pressure effects can be described by a shift 
of the jamming density $\varphi_J(P)$ and the friction
coefficient $\mu_J(P)$. Empirically, we find that these deviations 
from the hard sphere values $\varphi_J(P \to 0)$ and 
$\mu_J(P \to 0)$ are well described by the following formula:
\begin{eqnarray}
\varphi_J(P) & =\varphi_J(0)+c_\varphi P^{x},
\label{phi_P} \\
\mu_J(P) & =\mu_J(0)-c_\mu P^{y},
\label{mu_P}
\end{eqnarray}
The numerical values of the two new exponents entering
Eqs.~\eqref{phi_P} and \eqref{mu_P} are $x \approx 0.75$ and $y
\approx 0.56$, obtained by fitting over the entire pressure 
range. The fits are shown as lines in Figs.~\ref{fig8}-c,d. 
The precise value of these exponents depend on the 
chosen repulsive force (with $\alpha_r=2$) between the soft 
particles. 

The extension of the traditional 
granular rheology to soft particles is based on the simple 
physical idea that soft particles behave very much as 
hard particles, but with a `renormalized' particle 
diameter and friction coefficient. Therefore, 
this approach is similar to earlier attempts to 
describe the rheology~\cite{OlssonTeitel2012} and thermal
dynamics~\cite{tom} of soft particles in terms of the corresponding
hard sphere system.

The soft granular rheology in Eqs.~(\ref{phi_scale}, \ref{mu_scale})
suggests that rheological measurements {\it performed at constant 
pressure} (rather than constant density) would yield results essentially 
indistinguishable from zero pressures. In other words, detecting 
softness effects seems challenging in a traditional
granular experimental setting. In particular, the viscosity of 
a non-Brownian suspensions of soft particles measured at constant 
pressure would diverge with the same power law as for hard grains, 
but at a slightly higher packing fraction. 

The main new effect introduced by the particle softness is 
the possibility to explore states which are above the jamming transition, 
thus allowing a description of the constant density 
rheology of the solid phase above $\phi_J$. 
In fact, by taking the zero shear rate limit in 
Eqs.~(\ref{phi_scale}, \ref{mu_scale}) we find that for finite
$P$, the volume fraction converges to a value larger than 
$\phi_J(P \to 0)$ (yielding a compressed packing of soft particles)
supporting a {\it finite yield stress}, $\sigma_Y$. 
From Eqs.~(\ref{phi_P}, \ref{mu_P}), we obtain the following 
analytic form: 
\be
\sigma_Y = \left[ \frac{\phi-\phi_J(0)}{c_\phi} \right]^{\frac{1}{x}}
\left[ \mu_J(0) - c_\mu \left( \frac{\phi-\phi_J(0)}{c_\phi}  
\right)^{\frac{y}{x}}  \right].
\label{yield}
\ee 
This expression shows that the yield stress vanishes asymptotically as 
a power law as the jamming density is approached, 
$\sigma_Y \sim (\phi - \phi_J)^{\alpha}$, where $\alpha = 1/x \approx 1.33$.
However, much as for the shear viscosity, we find that the yield stress
is actually the sum of two distinct power laws with exponents 
$1/x$ and $(1+y)/x$, the latter being asymptotically sub-dominant. 
Values of the yield stress exponents in the range $\alpha = 1.1 - 1.5$ 
have been reported before~\cite{teitel,recent_teitel,brian,hatano2}.

The soft granular rheology in Eqs.~(\ref{phi_scale}, \ref{mu_scale}) 
allows us to describe the rheology of non-Brownian soft particles
on both sides of the jamming transition accounting 
at once for a diverging viscosity below $\phi_J$ and the emergence 
of a finite yield stress above $\phi_J$, with non-trivial 
shear-thinning regimes at larger shear rates. A clear advantage 
of the present description
is that it naturally incorporates the rheology of hard non-Brownian
spheres as a well-defined reference point, which is 
automatically recovered
in the limit of infinitely hard particles. 

\subsection{Comparison with the original Olsson and Teitel analysis}
\label{sec:olsson}

We now compare the soft granular rheology in 
Eqs.~(\ref{phi_scale}, \ref{mu_scale}) to previous work.
A qualitatively similar scaling analysis of the rheology 
of soft particles was proposed by Olsson and Teitel in Ref.~\cite{teitel}.
It has then often been used to organize numerical and experimental 
data obtained in non-Brownian 
soft suspensions~\cite{teitel,OlssonTeitel2012,nordstrom,jose}. 

This approach was motivated by an idea similar to the one 
used in the previous section, namely an extension of the hard particle 
limit to soft particles. The Olsson-Teitel analysis 
relies on essentially three assumptions~\cite{footnote}:
(i) the viscosity of hard particles diverges 
as a power law when the volume fraction 
approaches the jamming transition density: $\eta \propto 
(\phi_J - \phi)^{-\beta}$;
(ii) a finite yield stress is obtained above the jamming
transition which increases as a power law: $\sigma_Y \propto 
(\phi - \phi_J)^\alpha$. 
(iii) these two regimes can be connected by scaling.
Mathematically, these assumptions yield the following model for the 
critical properties of soft particles near jamming:
\be
\sigma_{xy} = |\phi-\phi_J|^\alpha {\cal G}_{\pm} 
\left( \frac{\gdot}{|\phi-\phi_J|^{\alpha+\beta}}
\right),  
\label{teitel}
\ee
with the following asymptotic forms for the scaling functions:
${\cal G}_{+} (x \to 0) \sim const.$ (emergence of a yield stress),
${\cal G}_{-} (x \to 0) \sim x$ (Newtonian regime with
diverging viscosity), 
${\cal G}_{\pm} (x \to \infty) \sim x^{\alpha/(\alpha+\beta)}$
(emergence of a critical shear-thinning regime). 
This mathematical model implies that exactly at the jamming 
transition, $\phi = \phi_J$,  
the rheology is described by a nontrivial power law, 
$\sigma \sim \gdot^{\alpha/(\alpha+\beta)}$ corresponding to 
shear-thinning behavior. This model is appealing as it also
directly connects to the well-known (empirical) Herschel-Bulkley
rheology~\cite{wagner,OlssonTeitel2012}. 
The scaling form in Eq.~(\ref{teitel}) therefore makes the simple 
assumption that the three regimes (Newtonian, yield stress, 
shear-thinning) are connected by smooth functions. 

\begin{figure}
\psfig{file=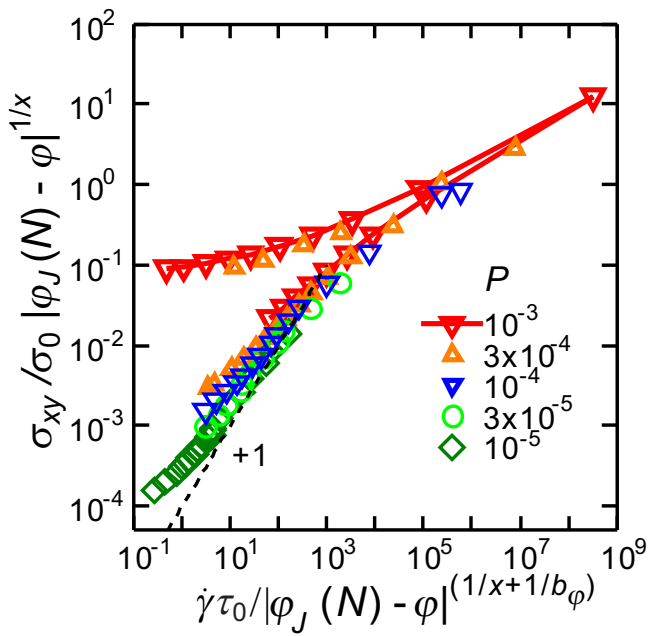,width=8.5cm,clip}
\caption{\label{fig9} Test of the scaling plot Eq.~(\ref{teitel}) for 
$N=10^4$ with exponents 
$\alpha = 1/x \approx 1.33$ from Eq.~(\ref{yield}) 
and $\beta=1/b_{\varphi} \approx 2.55$ from Eq.~(\ref{eta}). 
Deviations from perfect collapse are observed.}
\end{figure}

While very similar in spirit, the soft granular rheology 
we propose in Eqs.~(\ref{phi_scale}, \ref{mu_scale})
yields a model which is mathematically different from 
Eq.~(\ref{teitel}), although 
some asymptotic behaviors are the same. For instance, 
both models predict an asymptotic power law divergence of the 
Newtonian viscosity, power law emergence of the yield stress,
and algebraic critical shear-thinning regime. 
However, as we already pointed out, we find that  
both the Newtonian viscosity and the yield stress are
described by the sum of two power laws, 
see Eqs.~(\ref{eta}, \ref{yield}). 
A direct consequence of our model is that the 
rheology at the jamming density is described by a sum 
of three power laws, rather than a single one.
More importantly perhaps, all the {\it asymptotic behaviors
incorporated in our model cannot be simply connected by scaling 
functions} as in Eq.~(\ref{teitel}). 

We acknowledge that our conclusions are consistent with the more
recent description 
of the shear rheology near jamming by Olsson and Teitel themselves, which includes `corrections to 
scaling'~\cite{otherteitel2,note}. These corrections to scaling
are shown to affect the data collapse suggested by Eq.~(\ref{teitel}), 
and naturally predict that simple algebraic divergences become 
sums of power laws, in agreement with our findings. 
A similar statement is made 
more explicitely in other works~\cite{brian}, which also imply that 
the appealing data collapse predicted by Eq.~(\ref{teitel}) 
can only be approximate.  
In Fig.~\ref{fig9}, we confirm that our extended set of data cannot
be collapsed on two branches, using the scaling form
in Eq.~(\ref{teitel}). To build this diagram, we used 
asymptotic exponents for the yield stress and 
the shear viscosity determined from 
our simulations. We respectively use $\alpha=1/x$ obtained 
from Eq.~(\ref{yield}) and $\beta=1/b_{\varphi}$ obtained 
from Eq.~(\ref{eta}), together with our best estimate 
for the jamming density $\phi_J$. We have tried to 
use the freedom offered by Eq.~(\ref{teitel}) to shift these numbers 
from their actual values and used effective exponents and 
shifted critical density to improve the quality of the data collapse. 
We find that the data collapse cannot be significantly improved, 
in the sense that improving the quality of the collapse in one part 
of the plot deteriorates the quality in another part of the plot, and
acceptable collapse cannot be achieved for the entire data set.
These conclusions are also supported by direct analysis of the
mathematical model in Eqs.~(\ref{phi_scale}, \ref{mu_scale}) 
using numerical values 
determined from the simulations, where deviations from 
scaling can be analysed more finely.
We conclude that an apparent scaling of the data must result 
from the simultaneous use of a smaller dynamic range and 
of effective values for critical exponents.

\section{Summary and conclusions}
\label{sec:summary}

In this work we have optimized a simple numerical approach 
using non-Brownian soft particles to analyze in detail the 
rheology of non-Brownian suspensions over a very wide dynamic 
range, paying attention to finite size effects. 
This allowed us to confirm the algebraic divergence 
of the Newtonian viscosity of suspensions of hard particles and to
point out the difficulty of corresponding experimental 
measurements. Through a finite size scaling 
analysis, we have also established the existence of 
a correlation length scale that diverges as the jamming 
transition is approached.  

We then used these results to extend the precise hard sphere 
rheology obtained numerically to describe the non-linear rheology 
of soft particle suspensions across the jamming transition,
which we coined soft granular rheology. Although very simple
and natural from the viewpoint of granular materials, 
our approach yields a mathematical model which differs 
from earlier attempts at a scaling description. 
Because it starts from the natural `granular' variables 
$\mu$ and $\phi$, our approach suggests that no simple
scaling form exists for the shear viscosity of soft suspensions.

In future work we would like to explore in more detail 
the connection between the correlation length revealed in this work
and the non-local rheological effects which are currently 
receiving growing attention~\cite{KEP,kamrin,bruno}.
Another relevant issue concerns the role of frictional forces
in hard particle systems. It would be very interesting to 
extend the present study to include frictional forces, 
and see how our numerical results for the asymptotic behavior
of frictionless particles are affected by friction. This would
be very valuable to compare with experimental results 
performed with real granular suspensions where friction is 
unavoidable. 

\acknowledgments
We thank J.-N. Roux for useful discussions and S. Teitel and P. Olsson
for constructive comments on a previous version of this article.
The research leading to these results has received funding
from the European Research Council under the European Union's Seventh
Framework Programme (FP7/2007-2013) / ERC Grant agreement No 306845.

\end{document}